\documentclass[aps,preprint,showpacs,preprintnumbers,eqsecnum,amsmath,amssymb,11pt]{revtex4}

\usepackage{graphics,epsfig}
\usepackage{graphicx}
\usepackage{dcolumn}
\usepackage{bm}

\begin{document}

\title{Evolution of arbitrary spin fields in the
Schwarzschild-monopole spacetime}
\author{Qiyuan Pan \ \ \ \ Jiliang  Jing }
\email{Corresponding author, Email: jljing@hunnu.edu.cn}
\affiliation{ Institute of Physics and  Department of Physics, \\
Hunan Normal University,\\ Changsha, Hunan 410081, P. R. China }

\vspace*{0.2cm}
\begin{abstract}
\vspace*{0.2cm}

The quasinormal modes (QNMs) and the late-time behavior of arbitrary
spin fields are studied in the background of a Schwarzschild black
hole with a global monopole (SBHGM). It has been shown that the real
part of the QNMs for a SBHGM decreases as the symmetry breaking
scale parameter $H$ increases but imaginary part increases instead.
For large overtone number $n$, these QNMs become evenly spaced and
the spacing for the imaginary part equals to $-i(1-H)^{3/2}/(4M)$
which is dependent of $H$ but independent of the quantum number $l$.
It is surprisingly found that the late-time behavior is dominated by
an inverse power-law tail
$t^{-2[1+\sqrt{(s+1/2)^{2}+(l-s)(l+s+1)/(1-H)}]}$ for each $l$, and
as $H\rightarrow0$ it reduces to the Schwarzschild case
$t^{-(2l+3)}$ which is independent of the spin number $s$.

\end{abstract}

\vspace*{1.5cm}
 \pacs{03.65.Pm, 04.30.Nk, 04.70.Bw, 97.60.Lf}

\maketitle

The evolution of the external field perturbation around a black hole
is dominated by three successive stages
\cite{Frolov98,Kokkotas,Nollert}: the initial wave burst, the damped
oscillations called QNMs and the power-law tail behavior of the
waves at very late time. A well-known fact is that the QNMs have
become astrophysically significant with the realistic possibility of
gravitational wave detection in the near future because they are
entirely fixed by the structure of the background
\cite{vishe,Chand,Leaver,Cardoso,J-2}. In addition, the study of the
QNMs can also help us get a deeper understandings of the loop
quantum gravity \cite{Hod,Dreyer}, AdS/CFT and dS/CFT
correspondences \cite{Maldacena,Witten,Kalyana}. On the other hand,
the late-time evolution of various field perturbations has important
implications for two major aspects of black-hole physics: the
no-hair theorem and the mass-inflation scenario
\cite{Ruffini,Misner,Price,Leaver-1,Piran-tail,Hod-tail,LMBurko,J-4,J-5}.
Thus, we will discuss the evolution of massless arbitrary spin
fields around a SBHGM which is introduced by Barriola and Vilenkin
\cite{Barriola}, including the QNMs and late-time tails in this
short note. And our purpose is to see what effects the symmetry
breaking scale parameter and the spin number will have on the
evolution of the external field perturbation.

In the Boyer-Lindquist coordinates $(t, r, \theta, \varphi)$, the
metric for a SBHGM is \cite{Barriola,Bertrand,Salgado,Yu}
\begin{eqnarray}\label{metric}
ds^{2}=\frac{\Delta_{r}}{r^{2}}dt^{2}-\frac{r^{2}}{\Delta_{r}}dr^{2}
-r^2(1-H)(d\theta^{2}+sin^{2}\theta d\varphi^{2}),
\end{eqnarray}
where $\Delta_{r}=r^2-2M(1-H)^{-3/2}r$ and $H=8\pi\eta_{0}^{2}$, $M$
and $\eta_{0}$ represent the mass parameter of the black hole and
the symmetry breaking scale when the monopole is formed
respectively. Throughout this paper we use $G=c=1$. In this
spacetime the Teukolsky's master equations for massless arbitrary
spin fields \cite{Teukolsky-1,Teukolsky-2,Teukolsky-3,Teukolsky-4}
can be separated by using the Newman-Penrose formalism
\cite{Newman}. After the tedious calculation, we find that the
angular equation is the same as in the Schwarzschild black hole
\cite{Newman,Teukolsky-1,Teukolsky-2,Teukolsky-3,Teukolsky-4} and
the radial equation can be expressed as
\begin{eqnarray}\label{wave}
\frac{d^{2}\Psi_{s} }{d r_{*}^2}+[\omega ^2-V_{s}(r)]\Psi_{s} =0,
\end{eqnarray}
with
\begin{eqnarray}\label{Poten}
V_{s}(r)=is \omega r^2\frac{d}{d
r}\left(\frac{\Delta_{r}}{r^4}\right)+\frac{1}{r^4} \left[
\left(\frac{s}{2}\frac{d\Delta_{r}}{dr}\right)^{2}+
\left(s+\frac{\lambda}{1-H}\right)\Delta_{r}\right]
-\frac{\Delta_{r}}{r^3}\frac{d}{dr}\left[\Delta_{r}\frac{d}{dr}
\left(\frac{1}{r}\right)\right],
\end{eqnarray}
where $dr_{*}=(r^{2}/\Delta_{r})dr$ and
$\Psi_{s}=\Delta^{s/2}_{r}r^{1-2s}R_{s}$ ($R_{s}$ is the usual
radial wave function
\cite{Newman,Teukolsky-1,Teukolsky-2,Teukolsky-3,Teukolsky-4}),
$\lambda=(l-s)(l+s+1)$ is the angular separation constant for $s=0,\
\ -1/2,\ \ -1,\ \ -3/2$ and $-2$ with the quantum number
$l=|s|,|s|+1,\cdots$.

\textit{Quasinormal Modes} Introducing the expansion coefficients
$d_{n}$, we can give a solution to Eq. (\ref{wave})
\begin{eqnarray}\label{expand}
\Psi_{s}=r^{-\frac{s}{2}+2i\omega r_{+}}(r-r_{+})^{-\frac{s}{2}
-i\omega r_{+}}e^{i\omega r}\sum_{n=0}^{\infty}
d_{n}\left(\frac{r-r_{+}}{r}\right)^{n},
\end{eqnarray}
which satisfies the desired behavior at the boundary for the SBHGM
\cite{Chand}. Thus, the three-term continued fraction equation is
written as \cite{Leaver}
\begin{equation} \label{continued}
0=\beta_{0}-\frac{\alpha_{0}\gamma_{1}}
{\beta_{1}-}\frac{\alpha_{1}\gamma_{2}}
{\beta_{2}-}\frac{\alpha_2\gamma_3}
{\beta_{3}-}\frac{\alpha_3\gamma_4}{\beta_{4}-}\cdots,
\end{equation}
with
\begin{eqnarray}
&&\alpha_{n}=n^{2}+\left[2-s-\frac{4iM\omega}{\sqrt{(1-H)^{3}}}\right]n
+\left[1-s-\frac{4iM\omega}{\sqrt{(1-H)^{3}}}\right], \nonumber \\
&&\beta_{n}=-2n^{2}-\left[2-\frac{16iM\omega}{\sqrt{(1-H)^{3}}}\right]n
-\left[1+s-\frac{8iM\omega}{\sqrt{(1-H)^{3}}}
-\frac{32M^{2}\omega^{2}}{(1-H)^{3}}
+\frac{\lambda}{1-H}\right], \nonumber  \\
&&\gamma_{n}=n^{2}+\left[s-\frac{8iM\omega}{\sqrt{(1-H)^{3}}}\right]n
-\left[\frac{4isM\omega}{\sqrt{(1-H)^{3}}}
+\frac{16M^{2}\omega^{2}}{(1-H)^{3}}\right].
\end{eqnarray}
Eq. (\ref{continued}) has an infinite number of roots (corresponding
to the QNMs), but for each overtone number $n$, the QNMs
$\omega_{n}$ depend on the symmetry breaking scale parameter $H$,
spin number $s$ and quantum number $l$ if we take $2M=1$.

\begin{figure}[ht]
\includegraphics[scale=0.7]{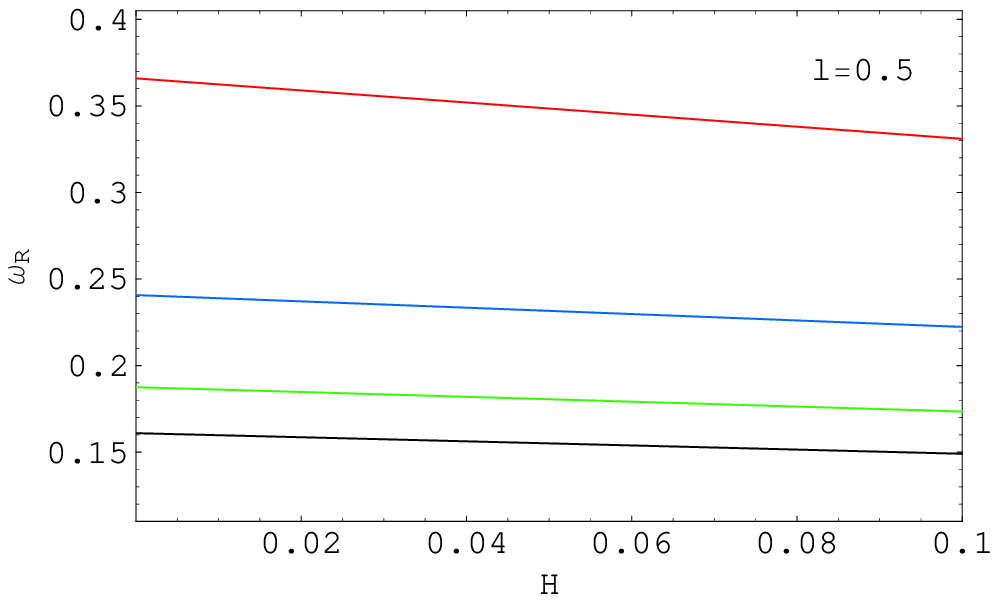}\hspace{0.2cm}%
\includegraphics[scale=0.7]{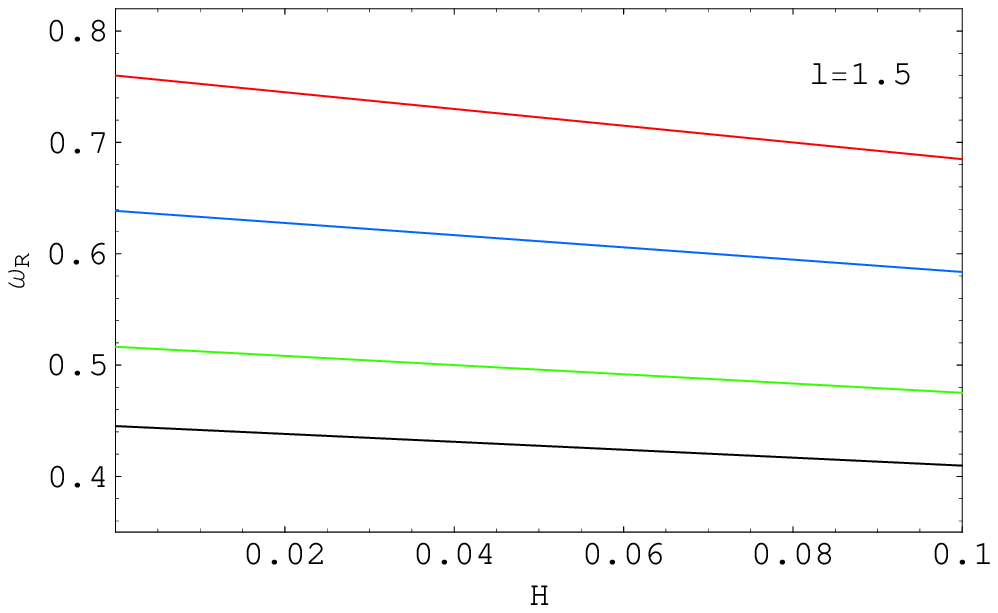}\\ \vspace{0.0cm}
\includegraphics[scale=0.7]{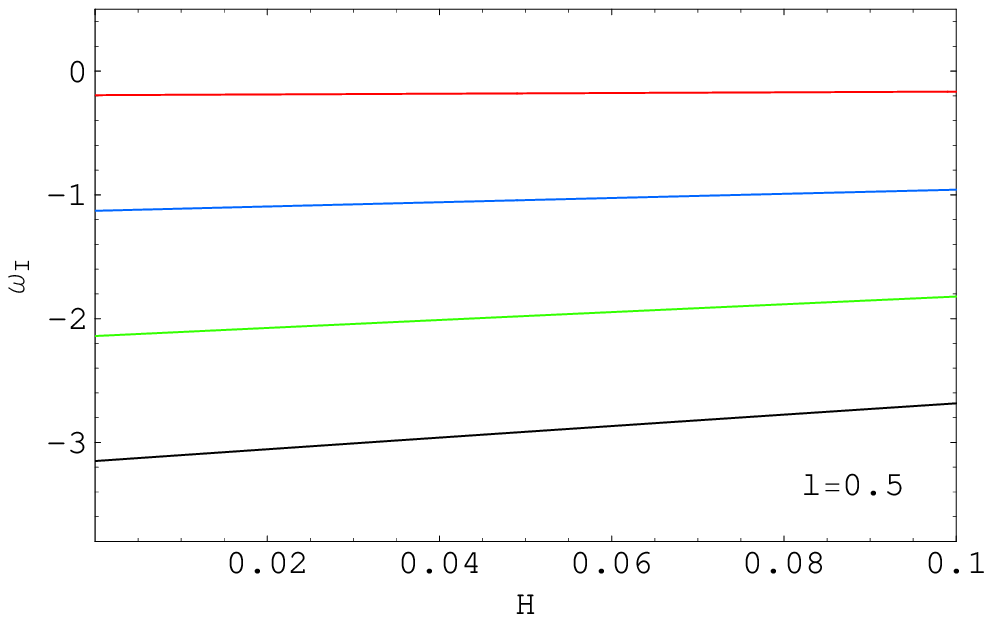}\hspace{0.2cm}%
\includegraphics[scale=0.7]{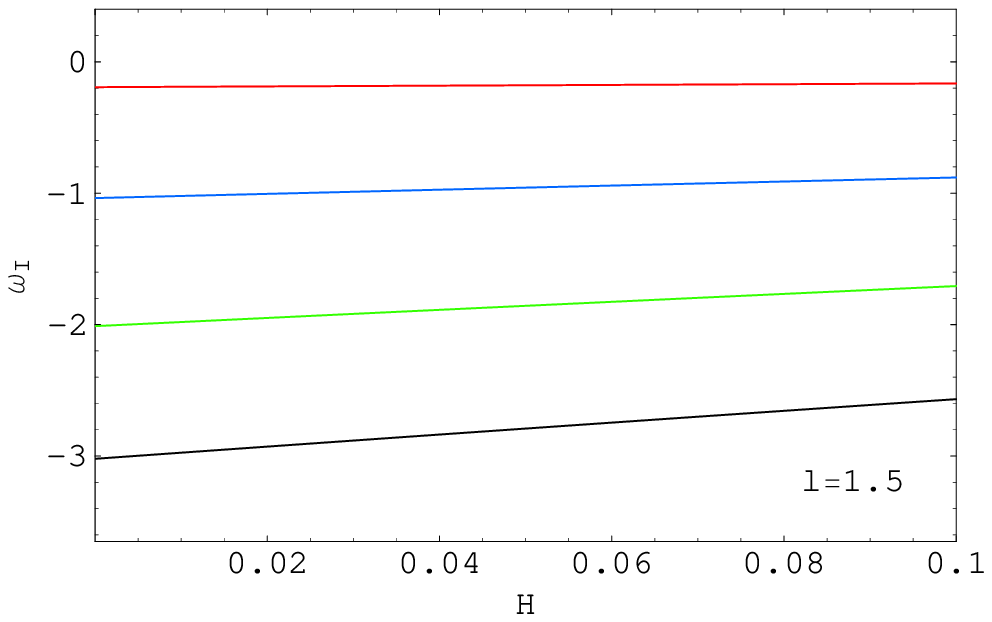}
\caption{\label{fig1}(Color online) Dependence of the real part
$\omega_{R}$ and imaginary part $\omega_{I}$ on $H$ for the Dirac
QNMs of the SBHGM. In each panel, the four lines from the top to the
bottom correspond to the modes for $n=0$ (red), $n=2$ (blue), $n=4$
(green) and $n=6$ (black) respectively.}
\end{figure}

\begin{figure}[ht]
\includegraphics[scale=0.7]{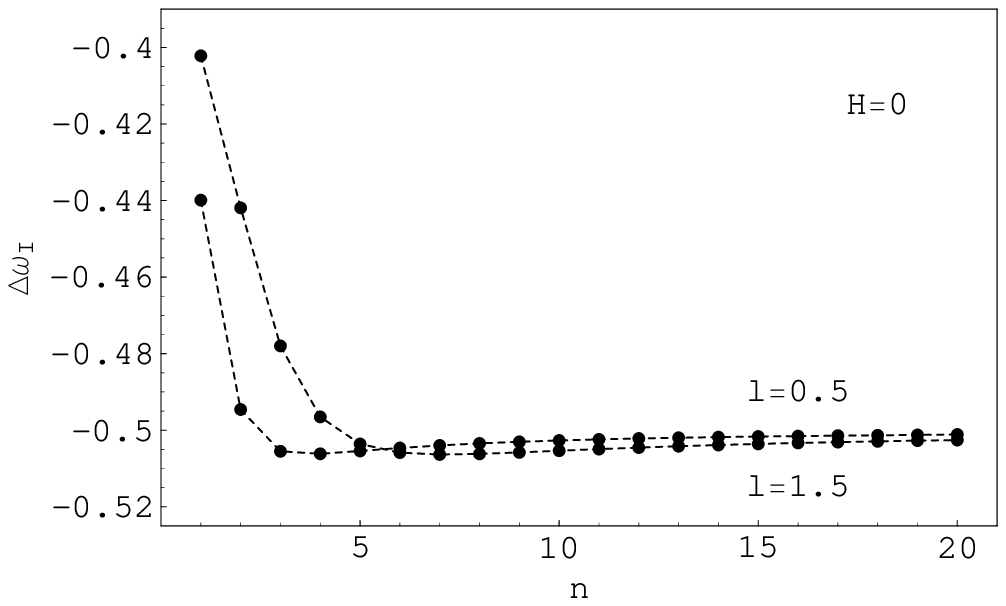}\hspace{0.2cm}%
\includegraphics[scale=0.7]{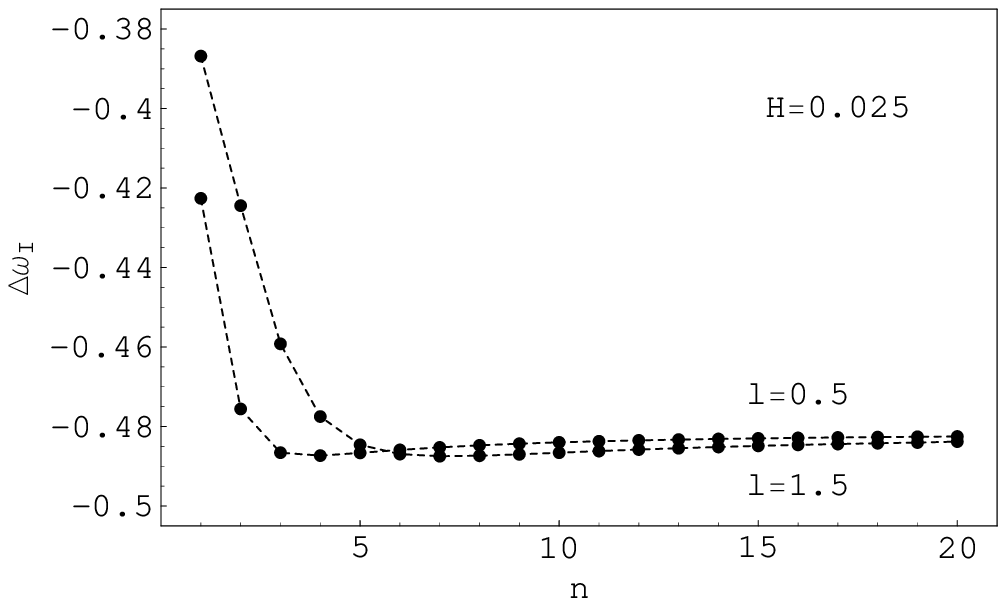}\\ \vspace{0.0cm}
\includegraphics[scale=0.7]{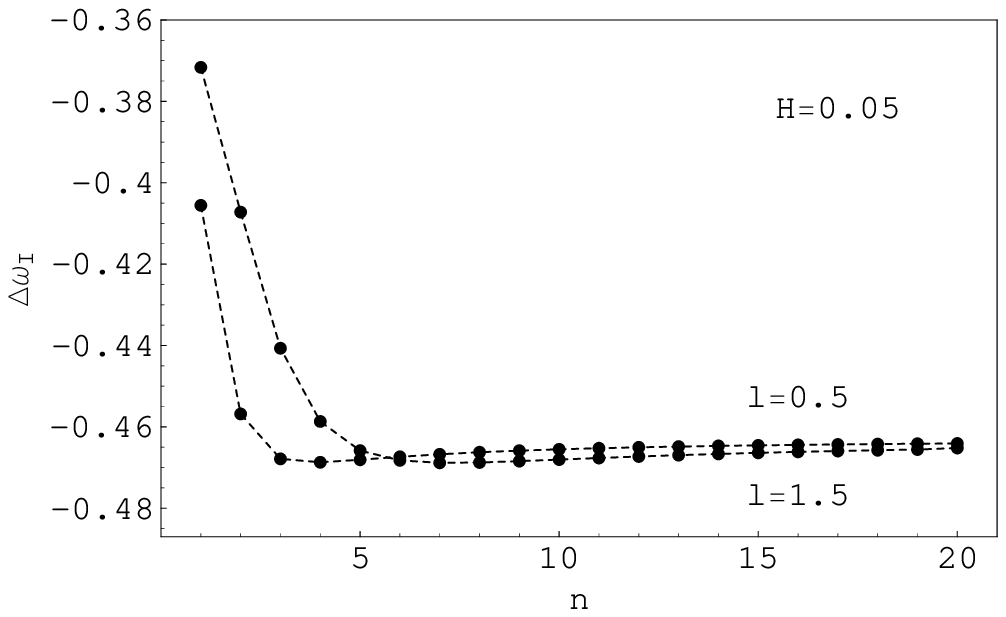}\hspace{0.2cm}%
\includegraphics[scale=0.7]{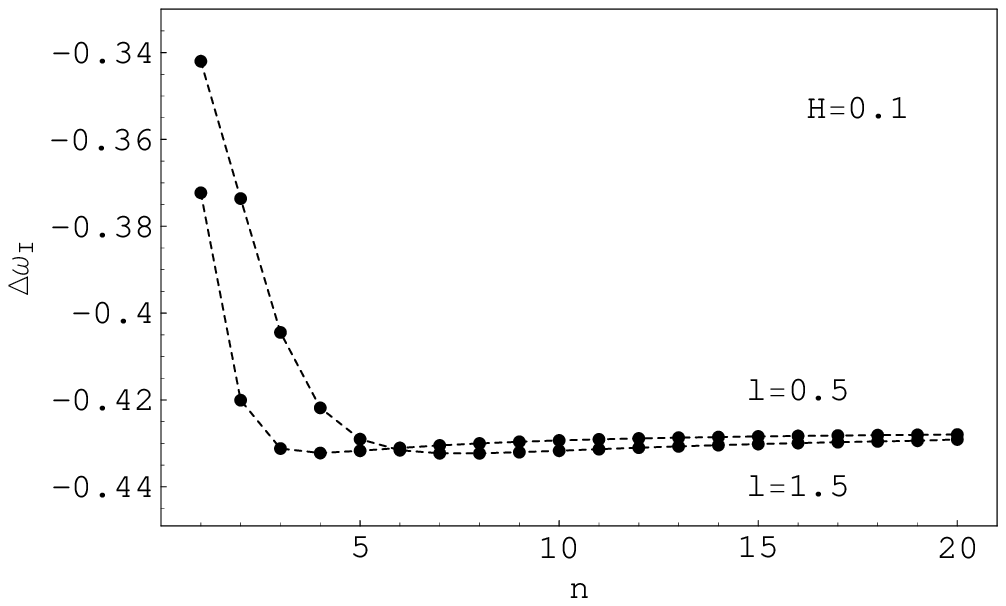}
\caption{\label{fig2}The spacing $\Delta\omega_{I}$ as the functions
of overtone number $n$ for $l=1/2$ and $l=3/2$. These panels shows
that $\Delta \omega_{I}$ is dependent of $H$ but independent of $l$
for large n.}
\end{figure}

For universality and clarity we only show some numerical results of
a massless Dirac field ($s=-1/2$) in Figs. \ref{fig1} and
\ref{fig2}. After discussing the other spin fields, we point out
that: (i) The QNMs of the massless arbitrary spin fields around the
SBHGM depend on $H$, and the real part $\omega_{R}$ of the QNMs
decreases as $H$ increases but imaginary part $\omega_{I}$
increases. (ii) The spacing for imaginary part of the QNMs for the
massless arbitrary spin fields in the background of the SBHGM is
\begin{eqnarray}
&&Im(\omega_{n+1})-Im(\omega_{n})\approx -
\frac{i}{4M}(1-H)^{3/2},~~~~{\text{as}}~~~n\rightarrow \infty,
\end{eqnarray}
which is dependent of $H$ but independent of $l$.

\textit{Late-time Behavior} Let us assume the observer and the
initial data are situated far away from the black hole. Then,
neglecting terms of order $0(\omega/r^{2})$ and the higher order
terms, we expand the wave equation (\ref{wave}) as a power series in
$M/r$
\begin{eqnarray}\label{Tail}
\left[\frac{d^{2}}{dr^{2}}+\omega^{2}+
\frac{2is\omega+4M(1-H)^{-3/2}\omega^{2}}{r}-\frac{1}{r^{2}}
\left(s+s^{2}+\frac{\lambda}{1-H}\right)\right] \xi_{s}(r,\omega)=0,
\end{eqnarray}
where $\xi_{s}=(\sqrt{\Delta_{r}}/r)\Psi_{s}$. So two basic
solutions required in order to build the Green's function can be
expressed as
\begin{eqnarray}
&&\tilde{\Psi}_{1}=Ar^{\rho+1/2}e^{i\omega r}
M(1/2+\rho+s-2iM(1-H)^{-3/2}\omega,1+2\rho,-2i\omega r),
\nonumber \\
&&\tilde{\Psi}_{2}=Br^{\rho+1/2}e^{i\omega r}
U(1/2+\rho+s-2iM(1-H)^{-3/2}\omega,1+2\rho,-2i\omega r),
\end{eqnarray}
where $\rho=[(s+1/2)^{2}+\lambda/(1-H)]^{1/2}$, $A$ and $B$ are
normalization constants. $M(a,b,z)$ and $U(a,b,z)$ are the two
standard solutions to the confluent hypergeometric equation
\cite{Abramowitz}. The function $U(a,b,z)$ is a many-valued
function, i.e., there will be a cut in $\tilde{\Psi}_{2}$.

It has been argued that the late-time tail is associated with the
existence of a branch cut (in $\tilde{\Psi}_{2}$)
\cite{Leaver-1,Piran-tail,Hod-tail}. The branch cut contribution to
the Green's function can be written as
\begin{eqnarray}
G^{c}_{s}(r_{*},r'_{*};t)=\frac{1}{2\pi}
\int_{0}^{-i\infty}\tilde{\Psi}_{1}(r'_{*},\omega)
\left[\frac{\tilde{\Psi}_{2}(r_{*},\omega e^{2\pi i})} {W_{s}(\omega
e^{2\pi i})}
-\frac{\tilde{\Psi}_{2}(r_{*},\omega)}{W_{s}(\omega)}\right]e^{-i\omega
t}d\omega,
\end{eqnarray}
where $W_{s}(\omega)=\tilde{\Psi}_{1}\tilde{\Psi}_{2,r_{*}}
-\tilde{\Psi}_{2}\tilde{\Psi}_{1,r_{*}}$ is the Wronskian. Since we
are only interested in the leading-order behavior at very late
times, we can assume that $\omega r'_{*}<\omega r_{*}<1$. Thus, with
the help of Ref. \cite{Abramowitz}, we obtain the late-time behavior
at timelike infinity
\begin{eqnarray}\label{TT}
G^{c}_{s}(r_{*},r'_{*};t)=\Upsilon(r_{*},r'_{*})~t^{-2(\rho+1)}
=\Upsilon(r_{*},r'_{*})~t^{-2[1+\sqrt{(s+1/2)^{2}+(l-s)(l+s+1)/(1-H)}]},
\end{eqnarray}
where $\Upsilon(r_{*},r'_{*})$ is the integral constant. To our
great surprise, this late-time behavior depends on the spin number
$s$, which is never found in the previous works. Taking $H=0$, we
get the late-time tails of the massless arbitrary spin fields for
the Schwarzschild black hole at timelike infinity, i.e.,
$t^{-(2l+3)}$ which is independent of the spin number $s$. This
result agrees with the Hod's analytical work for the massless
arbitrary spin fields on Kerr spacetimes \cite{Hod-tail}.

Summarizing, the QNMs of massless arbitrary spin fields in the
background of the SBHGM depend on the symmetry breaking scale
parameter $H$, and the real part of the QNMs decreases as $H$
increases but imaginary part increases instead. For large overtone
number $n$, these QNMs become evenly spaced and the spacing for the
imaginary part equals to $-i(1-H)^{3/2}/(4M)$ which is dependent of
$H$ but independent of the quantum number $l$. The most interesting
outcome of our study is that an inverse power-law tail
$t^{-2[1+\sqrt{(s+1/2)^{2}+(l-s)(l+s+1)/(1-H)}]}$ which depends on
$H$ and the spin number $s$ will dominate the late-time behavior,
which is shown that the SBHGM is quite different topologically from
a normal black hole due to the presence of a global monopole.

\begin{acknowledgments}
This work was supported by the National Natural Science Foundation
of China under Grant No. 10675045; the FANEDD under Grant No.
200317; and the SRFDP under Grant No. 20040542003.
\end{acknowledgments}

\end{document}